\documentclass[aps,amsfonts,amsmath,prd,tightenlines,preprint,showpacs,nofootinbib]{revtex4}
\usepackage{epsf}

\def\be{\begin{equation}}
\def\ee{\end{equation}}
\def\beq{\begin{equation}}
\def\eeq{\end{equation}}
\def\bea{\begin{eqnarray}}
\def\eea{\end{eqnarray}}
\def\bml{\begin{subequations}}
\def\blea{\bml\begin{eqnarray}}
\def\elea{\end{eqnarray}\end{subequations}}

\def\bx{\mathbf{x}}
\def\ba{\mathbf{a}}
\def\bb{\mathbf{b}}
\def\bp{\mathbf{p}}\def\phat{\hat\bp}
\def\bq{\mathbf{q}}\def\qhat{\hat\bq}

\begin{document}

\title{Cosmic string loops in the expanding universe}

\author{Ken D. Olum}
\email{kdo@cosmos.phy.tufts.edu}
\affiliation{Institute of
Cosmology, Department of Physics and Astronomy, 
Tufts University, Medford, Massachusetts  02155, USA}
\author{Vitaly Vanchurin}
\email{vitaly@cosmos.phy.tufts.edu}
\affiliation{Arnold-Sommerfeld-Center for Theoretical Physics,
Department f\"ur Physik, Ludwig-Maximilians-Universit\"at M\"unchen,
Theresienstr.\ 37, D-80333, Munich, Germany}

\begin{abstract}

We study the production of loops in the cosmic string network in the
expanding background by means of a numerical simulation exact in the
flat-spacetime limit and first-order in the expansion rate.  We find an
initial regime characterized by production of small loops at the scale
of the initial correlation length, but later we see the emergence of a
scaling regime of loop production.  This qualitatively agrees with
earlier expectations derived from the results of flat-spacetime
simulations.  In the final scaling regime we find that the
characteristic length of loops scales as $\sim 0.1 t$ in both
radiation and matter eras.

\end{abstract}


\maketitle

\section{Introduction}

Cosmological models with phase transitions naturally lead to the
production of topological defects, such as monopoles, cosmic strings,
and domain walls \cite{Kibble}.  Cosmic strings, in particular, can
assume a scaling solution in which the energy density in strings is a
fixed fraction of the total energy in matter or radiation, and this
network may exist today.  Recently, it was also realized that a
network of cosmic superstrings might be formed at the end of brane
inflation \cite{Tye,Dvali,Polchinski}.

In principle the cosmic strings could be observed directly by
gravitational lensing, and indirectly through CMB anisotropies and
gravitational wave background. However, the interpretation of any
observations and the setting of bounds from non-observation strongly
rely on knowledge of the distribution of energy in the network of
cosmic strings consisting of infinite strings and sub-horizon-size
loops.

It is very well established, both analytically
\cite{Kibble,Kibble:1984hp,Bennett:1986zn} and numerically
\cite{AT,BB,AS}, that the evolution of infinite strings exhibits
scaling. The interstring distance and the correlation length, scale
linearly with the horizon size (or, equivalently, the time).  Very
early numerical simulations \cite{AT} found loops at a large fraction
of the horizon size, but later work \cite{BB, AS} disagreed.  These
two groups found the loops were mainly produced at the smallest
resolution scale of the simulation, which did not scale.  This led to
the belief that such simulations, which treat the string as infinitely
thin and ignore gravitational back-reaction, would always give loop
production at the simulation resolution.  In a realistic cosmological
scenario, this behavior would presumably be cut off at a very small
scale by gravitational or perhaps even field-theoretic effects.
However, it is still not clear whether one should make such an
extrapolation starting from simulations that have dependence on the
resolution scale.

To overcome this problem, with Alex Vilenkin we recently developed an
exact simulation for the cosmic string network in flat spacetime
\cite{VOV}.  The precision of the simulation was only set by
unavoidable computer arithmetic uncertainty, which is many orders of
magnitude smaller than the scale of any structure in the network.
Using this code, we later showed that production of small loops was a
relic of the initial conditions in the simulation \cite{VOV2}. In the
long run the main production channel of loops was at the scales
comparable with the interstring distance $\sim 0.1 t$ . This
conclusion agrees with earlier theoretical predictions \cite{AV81,
Kibble85}.  We argued \cite{VOV2} that the stretching and smoothing
due to expansion would be unlikely to enhance the production of small
loops over that in flat spacetime, but the connection between
flat-spacetime and expanding-universe simulations is not completely
clear.

Recently two groups \cite{RSB,MS} did simulations with higher
resolution versions of the simulation codes previously used \cite{BB,
AS} and were able to resolve the loops.  Reference \cite{MS} found
some evidence for a scaling distribution of loops, but their size is
apparently much smaller than we \cite{VOV2} found; Ref.\ \cite{RSB}
found a distribution which diverges at small sizes and speculated that
the size would be cut off gravitationally.

In an attempt to settle this question, we here extend the simulations
of Ref.\ \cite{VOV2} to the expanding universe.  We are able to
simulate significantly later times than Refs.\ \cite{RSB,MS}.  We find
results very similar to that of Ref.\ \cite{VOV2}, with a peak in the
final production spectrum of loops at $l\sim 0.1t$.  Thus we expect
that the loop sizes in a real string network will be set by network
dynamics.  It is not necessary to invoke gravitational back-reaction
or finite string thickness to cut off the loop distribution at small
scales.

\section{Algorithm}

A cosmic string in flat spacetime obeys the Nambu-Goto equations of
motion, whose solution can be written
\be\label{eqn:Nambu}
\bx(t,\sigma)=\frac{1}{2}\left(\ba(\sigma-t)+\bb(\sigma+t)\right),
\ee
with constant functions $\ba(\sigma)$ and $\bb(\sigma)$ obeying
$\ba'^{2}=\bb'^{2}=1$.  By choosing piecewise linear functions $\ba$
and $\bb$, the world sheet of a string is represented by a set of flat
diamond-shaped regions in 4-space, and the simulation is exact (to the
limits of computer arithmetic).

In an expanding background, an exact solution is no longer possible.
However, we can form an approximate solution using Eq.\
(\ref{eqn:Nambu}) and adjusting $\ba$ and $\bb$ as the string evolves.
The approximation is good when the Hubble time is much larger than the
time for which $\ba'$ and $\bb'$ are held constant, which is also the
length of the segments of $\ba$ and $\bb$.  The simulation is then
exact in the limit of no expansion, and the corrections are first
order in the expansion rate.  The string is still represented in
(piecewise linear) functional form, and there is no minimum
resolution; arbitrarily small loops can be formed.

\begin{figure}
\begin{center}
\leavevmode\epsfxsize=3.0in\epsfbox{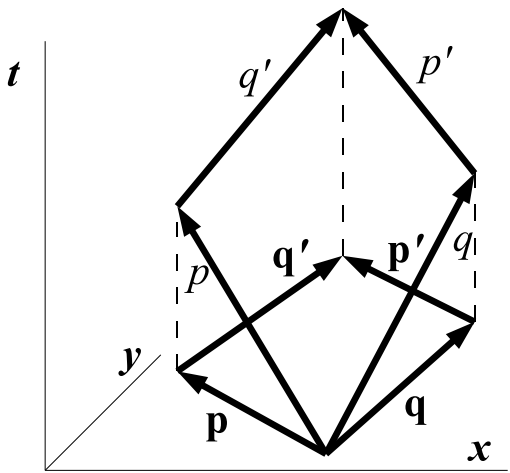}
\end{center}
\caption{A diamond-shaped region of a string world-sheet, and its
projection into a contant-time slice.  The vectors $\bp$ and $\bq$ are the
original left- and right-moving tangent vectors to the string.  The
change from $\bp$ and $\bq$ to $\bp'$ and $\bq'$ implements expansion
of the universe.}
\label{fig:diamond}
\end{figure}

We proceed as follows.  The coordinates are interpreted as comoving
spatial coordinates and conformal time.  We rename the unit vectors
\be
\phat\equiv -\ba',\; \qhat\equiv \bb'.
\ee
Since the universe is expanding, these vectors will not be constant.
The standard formulation of the equations of motion is \cite{BB} 
\be\label{eqn:pqcontinuous}
\dot{\phat}=-H \left[\qhat-(\phat\cdot\qhat)\phat\right],\quad
\dot{\qhat}=-H \left[\phat-(\phat\cdot\qhat)\qhat\right]
\ee
where $H =\dot a/a$ is the expansion rate, and the ratio of energy to
string parameter $\sigma$ is given by an auxiliary parameter
$\epsilon$, obeying
\be\label{eqn:epsiloncontinuous}
\dot\epsilon = -\epsilon H(1+\phat\cdot\qhat)
\ee
Rather than using unit vectors and keeping the parameter $\epsilon$
separate, we will take diamond-shaped regions of the string world
sheet spanned by vectors whose spatial parts are vectors $\bp =
p\phat$ and and $\bq = q\qhat$ in the past and new vectors $\bp'$ and
$\bq'$ in the future.  (See Fig.\ \ref{fig:diamond}.)
The lengths of these vectors should vary in proportion to $\epsilon$,
so
\be\label{eqn:pqdot}
\dot p/p = \dot q/q = \dot\epsilon/\epsilon
\ee
Using Eqs.\ (\ref{eqn:pqcontinuous}--\ref{eqn:pqdot}) we
can find the change in $\bp$ and $\bq$,
\be
\dot{\bp}=-H\left[p\qhat+\bp\right],\; \dot{\bq}= -H\left[q\phat+\bq\right]
\ee
Each point of the vector $\bp$ stays in the diamond for time
$q/2$, and so we find  the first-order expansions
\bea
\label{eqn:newplen}
p'&=&p-\frac{H}{2}\left[pq+\bp\cdot\bq\right] +O(H^2)\\
\label{eqn:newp}
\bp'&=&\bp-\frac{H}{2}\left[p\bq+q\bp\right] +O(H^2)
\eea
and similarly for $q$ and $\bq$.

This first-order analysis fixes the $O(H)$ terms but says nothing
about higher orders.  To fix the higher order terms, we proceed as
follows.  For the diamond to close, we require $\bp+\bq' =\bq+\bp'$.
Above $\bp$ and $\bq$ we construct null 4-vectors to be edges of the
diamond on the world sheet, as shown in Fig.\ \ref{fig:diamond}.  In
order for these two to intersect at a point, we require $p+q' = q+p'$,
which can be accomplished by setting the $O(H^2)$ term in Eq.\
(\ref{eqn:newplen}) to zero.  With $\bp'$ lying in the plane spanned
by $\bp$ and $\bq$, this fixes the $O(H^2)$ term in Eq.\
(\ref{eqn:newp}).  We treat $\bq'$ similarly.

Although all four vectors $\bp$, $\bq$, $\bp'$ and $\bq'$ lie on the
same plane (the $x$-$y$ plane in Fig. \ref{fig:diamond}), the
corresponding 4-vectors are null and therefore do not necessarily lie
in the same plane; in general the diamond is curved.  This is the most
important difference from a diamond-shaped region in flat spacetime.

The curved surface of a diamond in the expanding background is defined
by linear interpolation,
\be
\bx(\alpha,\beta) = \alpha\bp+\beta((1-\alpha)\bq+\alpha\bq')
= \beta\bq+\alpha((1-\beta)\bp+\beta\bp')
\ee
We then detect intercommutations as in Ref.\ \cite{VOV} by solving for
the places at which these curved surfaces intersect (which can be
reduced to solving a quartic equation) and perform the
intercommutations by adjusting the lists of $\bp$ and $\bq$ segments.

The above are the basic changes to implement the expansion of the
universe.  In addition, in the new context, a few changes to the
algorithm were necessary. The flat-spacetime simulation of Ref.\
\cite{VOV} kept track of the lengths of the loops at all times and
removed loops as soon as they were shorter than a cutoff.  Because
the segment lengths change due to expansion, this is not practical in
the expanding universe.  Instead we check periodically for loops which
have evolved for a complete oscillation without any intercommutations
and remove those from the simulation\footnote{It is not entirely
certain that such loops are in non-self-intersecting trajectories,
because of expansion, but it is an excellent approximation.}.  Because
of this change it is not possible to separate the primary production
spectrum of loops from the final spectrum after fragmentation, as was
done in Ref.\ \cite{VOV}.  Instead we give only the final spectrum.

For initial conditions we use a perturbed Vachaspati-Vilenkin (VV)
\cite{VV} procedure as in Ref.\ \cite{VOV}.  The initial VV cell size
is set to the initial Hubble distance.  The VV procedure is not
intended to precisely reproduce the result of a phase transition, but
merely to give strings which have no super-horizon correlations.
Since our goal is to study scaling, the ideal initial conditions would
already be in the scaling regime, but we do not know how to generate
those directly, so we use the VV procedure and let the evolution move
the network in the direction of scaling.  We do not use multiple
overlays as in Ref.\ \cite{VOV2}, in order to use less initial
computer memory and so allow a longer run without the increase of the
box size used in Ref.\ \cite{VOV2}.

In the present simulation,the perturbation procedure is improved over
previous work.  If one starts with a stationary network of VV strings
and decomposes each Nambu-Goto string into right- and left-moving
waves ($\ba$- and $\bb$-waves), then there are going to be preferred
cardinal directions of $\ba$- and $\bb$-waves, that are not relevant
to the physical picture. The preferred directions would remain
unchanged for the entire simulation in the flat spacetime and would
experience only small modifications in the expanding universe. In
fact, the distribution of $\ba'$ and $\bb'$ on the unit sphere is
sharply peaked around cardinal planes throughout such a
simulation. This problem remains if one simply adds random velocities
drawn from some normal distribution.  But a careful choice of random
velocities will allow a uniform distribution of $\ba'$ and $\bb'$ on
the unit sphere.  We divide the space of directions by drawing a
truncated rhombic dodecahedron around the origin.  If a segment of
string goes straight through a VV cube, we add a central point and
perturb it so the directions cover a square face of the dodecahedron,
and if it is curved in the cube we perturb it to cover half of one of
the hexagonal faces.

\section{Simulation time limit}

The simulation uses a periodic box of some size L, and one can trust
the results of such simulations only if the evolution is not affected
by the finiteness of the simulating box.  In principle, information
could travel through the box at the speed of light.  Two light rays
originating from the same spatial point and traveling in the opposite
cardinal directions would come together again at time
$L/2$.  Before that time the periodicity of the box cannot
matter.

In the string network the information is transmitted not by emitting
light rays, but through the strings themselves. Initially, the
strings are Brownian and in such strings the information can only
travel a distance of the order $\sqrt{T}$ by time $T$. However, the
evolution makes the strings smoother and the information can flow through
the network much faster.  

To be sure of the rate the information flows through the network, we
measure it directly from the simulation.  With each piece of string we
keep a rectangular region enclosing all points from which information
could possibly have affected the present location.  We define $s(t)$
as the maximum spatial extent along any cardinal direction of any
such region.  The information cannot travel faster than light, so the
function $s(t)$ cannot grow faster than $2t$. As long as $s(t)<L$, the
evolution of the network is not affected by the finite size of the
box, so we can set the running time of the simulation $T$ to
$s^{-1}(L)$.

\begin{figure}
\begin{center}
\leavevmode\epsfxsize=3.0in\epsfbox{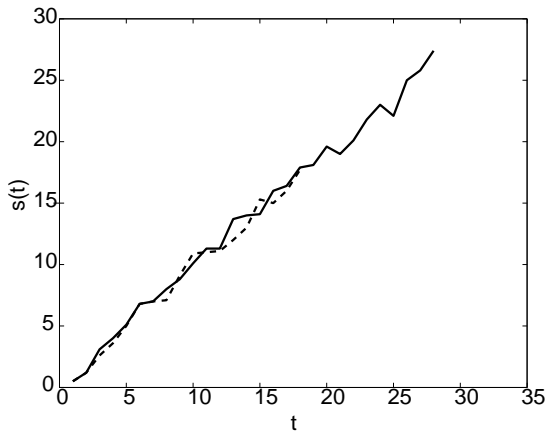}
\end{center}
\caption{Maximum total range $s(t)$ to which information travels for
box sizes 20 (dashed) and 30 (solid).  Since $s(t)\sim t$, one can
safely run the simulation until $t\sim L$.}
\label{fig:information}
\end{figure}

The function $s(t)$ extracted from two runs of box size 20
and of box size 30 is plotted in Fig.\ \ref{fig:information}
.
We see that $s(t)\sim t$, so that the maximum running time to be
certain there is no finite box effect should be of the order of the
box size: $T\sim L$.  Of course it is possible that the effects in
question are negligible and we could run for longer, but for the
present paper we will be conservative and stop our simulation at $t = L$.

Our simulations are done with box size 120 in units of the initial VV
cell size.  In the radiation era, we start from conformal time 1, so
our dynamic range is 120.  In the matter era we use the same box and
initial conditions, but the expansion is faster and the simulation is
too inaccurate at time 1, so we start from conformal time 2 and thus
have dynamic range 60.

\section{Results}

As in Ref.\ \cite{VOV2}, we characterize the rate of loop production
by the function $n(l, t)$ --- the number of loops produced per unit
loop length per unit volume of the network per unit time.  Here we
define a loop to be ``produced'' at the time when it enters a
non-self-intersecting trajectory.  Thus if a loop is produced from the
network and later fragments, we do not count the original loop but
only the final fragments in the production function.  

\begin{figure}
\begin{center}
\leavevmode\epsfxsize=3.0in\epsfbox{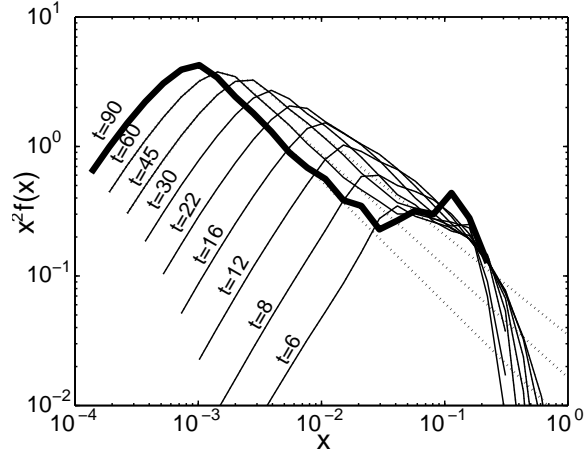}
\end{center}
\caption{The production rate of loops in the radiation era.  Each
graph is the loop production rate averaged around the labeled time.
The vertical axis is $x^2 f(x)$, which is the rate of length emitted
into loops per logarithmic interval of $x$.  Dotted lines are
power-law extrapolations of the right side of the first peak, for times
(from top to bottom) 45, 60, and 90.}
\label{fig:radiation}
\end{figure}

\begin{figure}
\begin{center}
\leavevmode\epsfxsize=3.0in\epsfbox{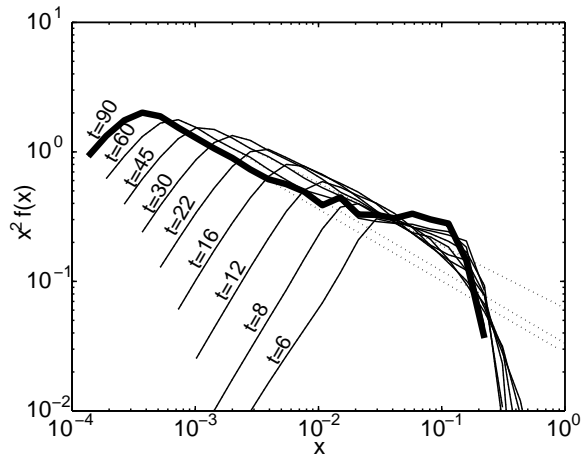}
\end{center}
\caption{As in Fig.\ \ref{fig:radiation} for the matter era.}
\label{fig:matter}
\end{figure}

In a scaling network \cite{VOV2},
\be
n(l, t) = t^{-5} f(x),
\ee
where $f$ can be any function of $x = l/t$.  The function $f(x)$ is
plotted in Fig.\ \ref{fig:radiation}
for the radiation era and in Fig.\ \ref{fig:matter}
for the matter era. Unlike in the flat-spacetime simulation, where loops
smaller than some fraction of the horizon are removed, no artificial
cut-off is present in the expanding-universe simulations. String loops
are only removed when they are in a non-self-intersecting
trajectory. By that time their physical size is much smaller than the
interstring distance and they are not likely to interact with the
rest of the network.

Scaling in loop production means production of loops at a fixed
fraction of the horizon size, and thus a fixed $x$.  So a scaling
feature in the loop production spectrum is one which appears at a
fixed horizontal position in Figs.\ \ref{fig:radiation} and
\ref{fig:matter}.  Indeed such a feature exists in these figures.  At
late times a peak appears around $x = 0.1$.  There is also a (higher)
non-scaling peak, which moves to the left as time passes.  We believe
that this peak is due to the initial conditions, which are not exactly
the conditions of a scaling network.  In the long run, when the
structures on the scales of the initial condition are smoothed out, we
expect the non-scaling peak to die away and the scaling
peak to dominate the evolution.

\begin{figure}
\begin{center}
\leavevmode\epsfxsize=3.0in\epsfbox{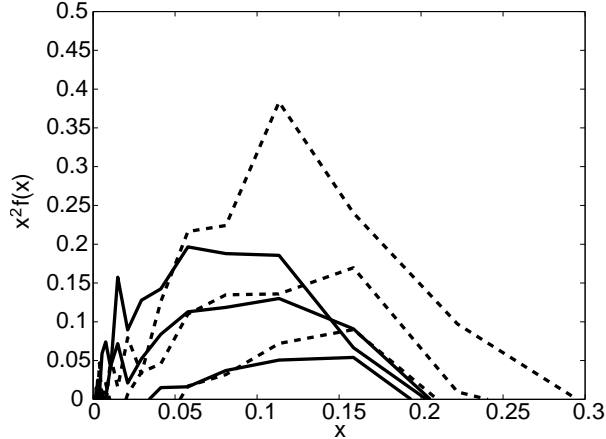}
\end{center}
\caption{The scaling peak of the loop production rate with the
extrapolated non-scaling peak subtracted, for times (from bottom to
top) 45, 60, and 90.  The radiation era is shown dashed and the matter
era solid.  The scaling peak in both cases is very approximately at
$x=0.1$.}
\label{fig:scaling}
\end{figure}

The scaling behavior can be seen better if one excludes the loops
contributing to the non-scaling peak.  We make a power law
extrapolation of the non-scaling contribution (presumably an
overestimate) to larger values of $x$, shown in Figs.\
\ref{fig:radiation} and \ref{fig:matter} with dotted lines.
Subtracting this extrapolated contribution gives a better view of the
scaling peak, shown in Fig.\ \ref{fig:scaling}.

It also worth mentioning that our expansion algorithm becomes more and
more exact with time, since $H$ becomes smaller and smaller while the
segments of string do not get longer.  Therefore, the growth of the
scaling peak at late times cannot be attributed to the inaccuracy of
the first-order approximation.

\section{Conclusion}

We have presented the results of expanding-universe simulations of a
cosmic string network.  Our simulations keep the string position and
velocity in functional form and have no minimum resolution.  They are
exact in the flat-spacetime limit; in the expanding universe they use an
approximation accurate to first order in the ratio of the string
segment size (from the initial conditions) to the Hubble distance.

The loop production is originally dominated by loops whose size results
from the initial conditions, but at late times there is a new, scaling
peak in the loop production spectrum.  The peak value of loop length
is about $0.1 t$ in flat spacetime, in the radiation era, and in the
matter era.

Our simulations run until $t=L$, where $t$ is conformal time and $L$
is the box size, in this case 120 times the initial Vachaspati-Vilenkin cube
size.  We start at $t = 1$ in the radiation era, so the ``dynamic
range'' (final/initial time) is 120, much longer than in other
simulations (3 in Ref.\ \cite{MS}, 8 in Ref.\ \cite{RSB}).  Because of
this difference, we are able to see the late time behavior of the
string network that is quite different from the initial evolution.

\section*{Acknowledgments}

We are grateful to Alex Vilenkin for suggesting and helping to develop
the original exact simulation technique and for useful advice in the
present project, and to Paul Shellard for helpful discussions. This
work was supported in part by the National Science Foundation under
grants 0353314 and 0457456, and by project ``Transregio (Dark
Universe)''.

\end{document}